\documentclass[aps,showpacs,twocolumn,superscriptaddress,pra]{revtex4}

\usepackage{amsmath}
\usepackage{amssymb}
\usepackage{amscd}
\usepackage{wasysym}
\usepackage[ansinew]{inputenc}
\usepackage[T1]{fontenc}
\usepackage{ae,aecompl}

\usepackage[pdftex]{graphicx}
\DeclareGraphicsExtensions{.pdf}
\newcommand{\be}{\begin{equation}}
\newcommand{\ee}{\end{equation}}
\newcommand{\bea}{\begin{eqnarray}}
\newcommand{\eea}{\end{eqnarray}}
\newcommand{\ba}{\begin{align}}
\newcommand{\ea}{\end{align}}
\newcommand{\nn}{\nonumber}

\newcommand{\ket}[1]{|#1\rangle}
\newcommand{\bra}[1]{\left\langle#1\right|}

\newcommand{\ri}{{\rm i}}
\newcommand{\rd}{{\rm d}}
\newcommand{\upd}{{\rm d}}

\begin{document}

\bibliographystyle{apsrev}
\title{Photon sorters and QND detectors using single photon emitters}
\author{D. Witthaut}
\affiliation{QUANTOP, Niels Bohr Institute, University of Copenhagen, DK--2100 Copenhagen \O{}, Denmark}
\affiliation{Max-Planck-Institute for Dynamics and Self-Organization (MPIDS),
D--37077 G\"ottingen, Germany}
\author{M. D. Lukin}
\affiliation{Department of Physics, Harvard University, Cambridge MA 02138}
\author{A. S. S\o{}rensen}
\affiliation{QUANTOP, Niels Bohr Institute, University of Copenhagen, DK--2100 Copenhagen \O{}, Denmark}
\date{\today }

\begin{abstract}
We discuss a new method for  realizing number-resolving and non-demolition  photo detectors by  strong coupling of light to individual single photon emitters, which act as  strong optical non-linearities. As a specific application we show how these elements can be integrated into an  error-proof Bell state analyzer, whose efficiency exceeds  the best possible performance with linear optics even for a modest atom-field coupling. The methods are  error-proof in the sense that every detection event unambiguously  projects the photon state onto a Fock or Bell state and imperfections only result in reduced success probability, not in wrong results.
\end{abstract}

\maketitle


\section{Introduction}

Experimental realization of number-resolving,  non-demolition photo 
(QND) detectors is a long-standing challenge in quantum optics and 
quantum information science. 
{Progress has been made in the microwave regime {\cite{Nogu99,Guer07,Schu07,John10}}, but the optical 
regime remains an unsolved challenge.}
Conventional  photodetectors measure 
only the  intensity or the energy of an incoming light pulse, and are 
not capable of measuring photon states in a QND fashion. More 
advanced measurements schemes can be constructed using 
optical non-linearities, but these are typically very weak since 
photons rarely interact with each other. In this Letter, we show 
how  to overcome this problem by exploiting strong coupling of light to 
individual single photon emitters. This  provides a strong optical 
non-linearity {(cf.\cite{Chan07b})}, which enables the realization 
of number-resolving photon sorters and quantum non-demolition 
photo detectors.

A common approach to realizing strong coupling between
photons and {emitters relies on cavity quantum 
electrodynamics (QED), where the light field is confined to a 
high--Q optical resonator \cite{Thom92,Duan04}
or a microcavity \cite{Reit04,Yosh04,Daya08}}. 
An alternative approach is to use single emitters coupled to one 
dimensional photonic waveguides and great advances have been 
made {using tapered optical fibers coupled to a single atom 
\cite{Vets10}}, microwave transmission lines 
coupled to a flux qubit \cite{Asta10}, or surface plasmons modes
coupled to a single photon emitter \cite{Chan06,Akim07,Chan07b,Huck11}.
In these systems the emitter couples to a continuous one-dimensional 
spectrum of modes and photon scattering is governed by the 
interference of absorbed, reemitted, and directly transmitted waves 
\cite{Shen05,Shen07a,Shen07b,10scatter3}. 
In a similar way, the transmission of a tighly focussed light
beam can be controlled by a single emitter in free space 
\cite{Hwan09,Stob09}.
In the present paper we will explore possible applications of emitters 
coupled to such a one-dimensional photonic continuum for photo 
detection, but the ideas and formalism we use can also be applied to 
cavity QED as well as other methods of achieving strong optical 
non-linearities {\cite{Daya08,Imam97b,Luki00,Andr05}}.

First, we consider passive devices based on simple two-level
emitters. The interaction with the emitter naturally leads to a
photon sorter effect, which can be used to implement a 
number resolving photo detector.
Secondly, we consider a waveguide coupled to a three-level 
emitter controlled by a classical laser field. This setup offers 
significantly more opportunities at the expense of a more 
complex setup. In particular we discuss QND photo detectors. 
As a possible {application}
of these devices we will show how to construct optical Bell-state 
analyzers. 
A Bell measurement is an essential ingredient in quantum 
information, as it enables efficient quantum repeaters \cite{Duan01}
as well as universal optical quantum computers \cite{Knil00}. 
Unfortunately such a measurement cannot be realized with 
linear optics \cite{Luet99}, but requires a strong nonlinearity. 
We focus on realistic systems with losses and discuss how to make
devices error-proof. Even in the case of an error, a measurement  
shall at most give an inconclusive but never a wrong result. 
This is important for many devices in quantum communication,
in particular quantum repeaters \cite{Duan01},
where incorrect measurement outcomes spoil the results whereas 
inconclusive outcomes only increase the necessary resources.

\section{Waveguide quantum optics}

To begin with, we briefly review the theory of photon scattering
in a (semi-) infinite single-mode waveguide introduced in 
Refs. \cite{Shen05,Shen07a,Shen07b,10scatter3}.  
Following this approach, the photon field is decomposed into 
a left- and a right-moving mode described by the bosonic field 
operators $\hat a_L(x)$ and $\hat a_R(x)$, respectively. 
We assume that the photons have a linear dispersion relation with
propagation velocity $c$ for the entire frequency range of interest.
The light-field couples to a single emitter placed inside the 
waveguide or in the evanescent field.

\begin{figure}[tb]
\centering
\includegraphics[width=8cm, angle=0]{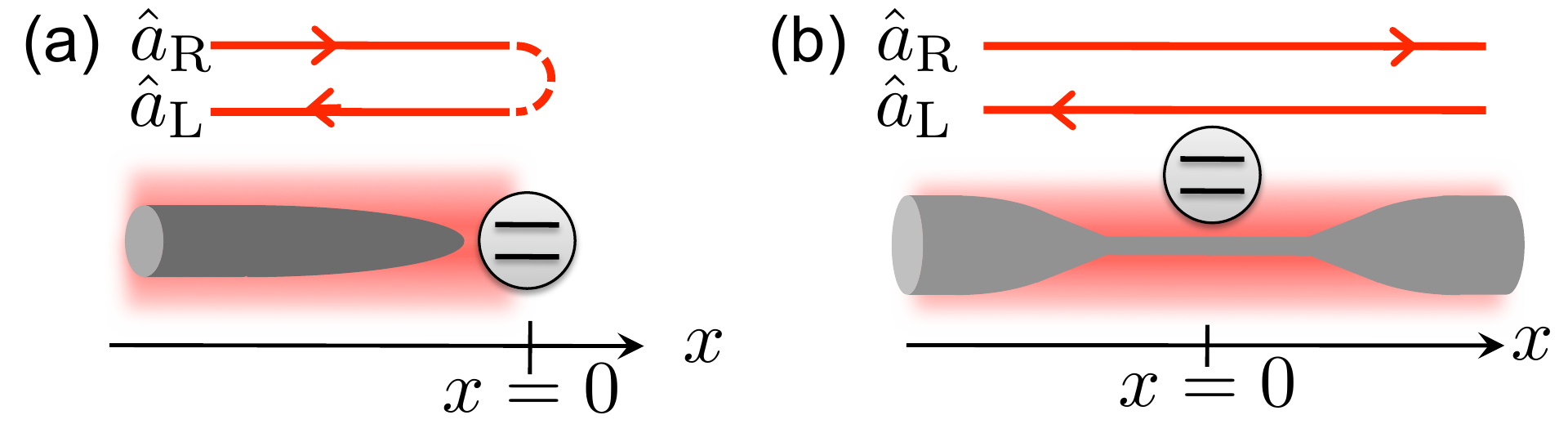}
\caption{\label{fig-exp}
Sketch of potential experimental setups illustrating the used
coordinate system and the naming of the optical modes.
(a) End-coupling as proposed for plasmonic waveguides \cite{Chan07b}.
(b) Side-coupling to a tapered optical fibers \cite{Vets10}.
}
\end{figure}

For some systems the most efficient coupling is realized by 
placing the emitter at one end of a semi-inifinite waveguide instead 
of side-to-side, in particular for plasmonic waveguides \cite{Chan07b}.
For these systems we take the position of the end of the guide to 
be $x=0$. The propagation of the right- and left-going modes is 
then restricted to $x<0$, {cf.~Fig.~\ref{fig-exp} (a)}. 
In this case we introduce the mode function
\be
  \hat a_e(x) = \left\{ \begin{array}{c l}
    \hat a_R(x) & \mbox{for} \; x < 0 \\
    \hat a_L(-x) & \mbox{for} \; x > 0, \\
    \end{array} \right.
\ee
such that $x<0$ describes the incoming and $x>0$ the reflected
photons. The free photon field is then described by the Hamiltonian
\be
  \hat H_{\rm free} = - \ri c \int \rd x \,
   \hat a_e^\dagger(x) \frac{\partial}{\partial x } \hat a_e(x).
\ee 

The coupling to the emitter is assumed to be local at the position 
of the end, $x=0$. Within the rotating wave approximation, 
photon scattering is then described by the Hamiltonian
\be
  \hat H = \hat H_{\rm atom} + \hat H_{\rm free}
    +  \sqrt{\Gamma} \int \rd x \, \delta(x) 
   \left( \hat S_-  \hat a_e^\dagger(x) +  \hat S_+  \hat a_e(x) \right),
   \label{eqn-ham1}
\ee
where $\hat S_+ = \hat S_-^\dagger$ is an atomic raising operator. 
We set $\hbar=1$, thus measuring all energies in frequency units. 
The coupling strength is here parametrized by the rate of  spontaneous emission 
into the waveguide $\Gamma$.

If the emitter is coupled to a waveguide extending in both directions, 
we have to keep both the left- and the right-moving modes $\hat a_L(x)$ 
and  $\hat a_R(x)$, {cf.~Fig.~\ref{fig-exp} (b)}. 
Nevertheless, the emitter only {couples} to the symmetric mode
$\hat a_e(x) =  (\hat a_R(x) + \hat a_L(-x))/\sqrt{2}$, again taking
the position of the emitter as zero. Thus we recover the Hamiltonian 
(\ref{eqn-ham1}) also in this setup.
Below we focus on the semi-infinite case, which is conceptually simpler. 
However, the same detectors can be realized for waveguides 
extending in both directions by directing any incident {photon} to
a balanced beam splitter to prepare a symmetric superposition 
of left and right moving modes.

The basic scattering problem for one photon interacting with a 
two-level emitter has been solved in \cite{Shen05}. In this case we 
have $\hat H_{\rm atom} =  (\omega_0-i\gamma/2) \ket{e}\bra{e}$
and $\hat S_+ = \ket{e}\bra{g}$, where $\ket{g}$ and $\ket{e}$ are 
the ground and excited state of the emitter, respectively. Similar to 
the quantum jump approach \cite{Dali92} we have added an imaginary part 
to the resonance frequency of the emitter $\omega_0 - i \gamma/2$ 
describing a coupling of the emitter to other modes than the waveguide 
with a rate $\gamma$. 
Using the Lippmann Schwinger formalism one can then show that a
monochromatic input state with wave number $k$ is scattered as
\be  
  \ket{\Psi_{\rm in}} = \hat a_k^\dagger \ket{0} 
  \quad \longrightarrow \quad
  \ket{\Psi_{\rm out}} = t_k \hat a_k^\dagger \ket{0}.
\ee  
Here, $\ket{0}$ denotes the vacuum state, i.e. the empty waveguide,
and the operator $\hat a_k^\dagger$, defined as the Fourier transform 
of $\hat a_e^\dagger(x)$, creates a photon with wave number $k$.
The transmission amplitude is given by \cite{Shen05}
\be
  t_k = \frac{ck - \omega_0 + \ri ( \gamma - \Gamma)/2}{
                   ck - \omega_0 + \ri ( \gamma + \Gamma)/2} \, .
  \label{eqn-tk}
\ee
If losses are negligible, $\gamma = 0$, the photon thus {experiences} 
only a phase shift which equals $\pi$ on resonance $ck = \omega_0$.
If the emitter is side-coupled to an infinitely extended waveguide,
an incident photon is reflected with a probability $|1-t_k|^2/4$, which 
can be shown by transforming back to left- and right-moving modes.
Scattering by a three-level emitter was discussed in detail in
ref.~\cite{10scatter3}. 

When two or more photons enter the waveguide simultaneously, 
scattering by a two-level emitter induces an effective photon-photon 
interaction, which is extremely useful for the detection and manipulation
of light fields.
Here we consider an input state of two identical photons 
with pulse shape $f_{2, \rm in}(k,p)  = f_1(k) f_1(p)$,
\be
  \ket{\Psi_\mathrm{in}} = \frac{1}{\sqrt{2}} \int \upd k \, \upd p \, 
    f_{2, \rm in}(k,p) \, \hat a_k^\dagger  \hat a_p^\dagger  \ket{0},
  \label{eqn-2phot-in}
\ee 
{where $k$ and $p$ denote the wave numbers of
the two photons.}
The interaction with a single two-level emitter introduces
{strong correlations of the two photons}
\be
  f_{2, \rm out}(k,p) = t_k t_p f_1(k) f_1(p) + f_B(k,p).
  \label{eqn-fbound}
\ee
{Here, the subscript $B$ refers to a 'bound state' contribution,
whose precise form can be found in Refs.~\cite{Shen07a,Shen07b}.}

Let us briefly comment on the applicability of the theoretical 
model and possible errors in an experimental realization.
The most obvious source of errors is the imperfect coupling of 
the emitter to the waveguide. This is taken into account by the 
coupling rate $\gamma$ to other modes than the waveguide.
Indeed, the success probability of all devices introduced in this 
letter crucially depends on the ratio $\Gamma/\gamma$.
Currently, the most advanced experimental setup employs a
microwave transmission line side-coupled to a flux qubit.
In this experiment, a photon incident from the side was back-scattered 
with 94\% probability. Translated into the current setting this 
corresponds to a coupling of $\Gamma/\gamma \approx 32$.
In the optical regime, a strongly increased decay rate was 
demonstrated for single quantum emitters coupled to a 
surface plasmon polariton mode on a silver nanowire \cite{Akim07}.
For a single diamond color center, an increase by a factor 
of 3.6 was reported \cite{Huck11}. This suggests (but does 
not prove) $\Gamma/\gamma \approx 2.6$.
The second major source of errors is photon absorption
in the waveguide. While this process is negligible for tapered optical 
fibers and free space setups \cite{Daya08,Vets10,Hwan09,Stob09}, 
losses are particularly strong in plasmonic systems. In such an 
experiment a rapid in- an outcoupling to a dielectric waveguide 
is inevitable (cf.~\cite{Chan07b}). However, we note that 
absorption is a general problem which is equally present in linear
photo detectors. In addition, all measurement schemes presented
in this letter are designed such that photon loss may lead to 
inconclusive, but not to incorrect results. In particular, a successful 
Bell state measurement always requires the coincidence detection 
of both incident photons for the setup presented below as well as 
for any setup based on linear optics. Hence inefficient detectors 
would suppress both linear optics and {the} more advanced 
detectors presented here by the same amount. Additional coupling 
losses can easily be accounted for by multiplying our resulting 
success probabilities by the probability for the photons not to be lost.
A less obvious, but more critical issue is the dispersion of properties 
from emitter to emitter, especially in a solid-state realization. This
source of errors is analyzed in detail below for the photon sorter device.

\section{Photon sorter}
\begin{figure}[tb]
\centering
\includegraphics[width=8cm, angle=0]{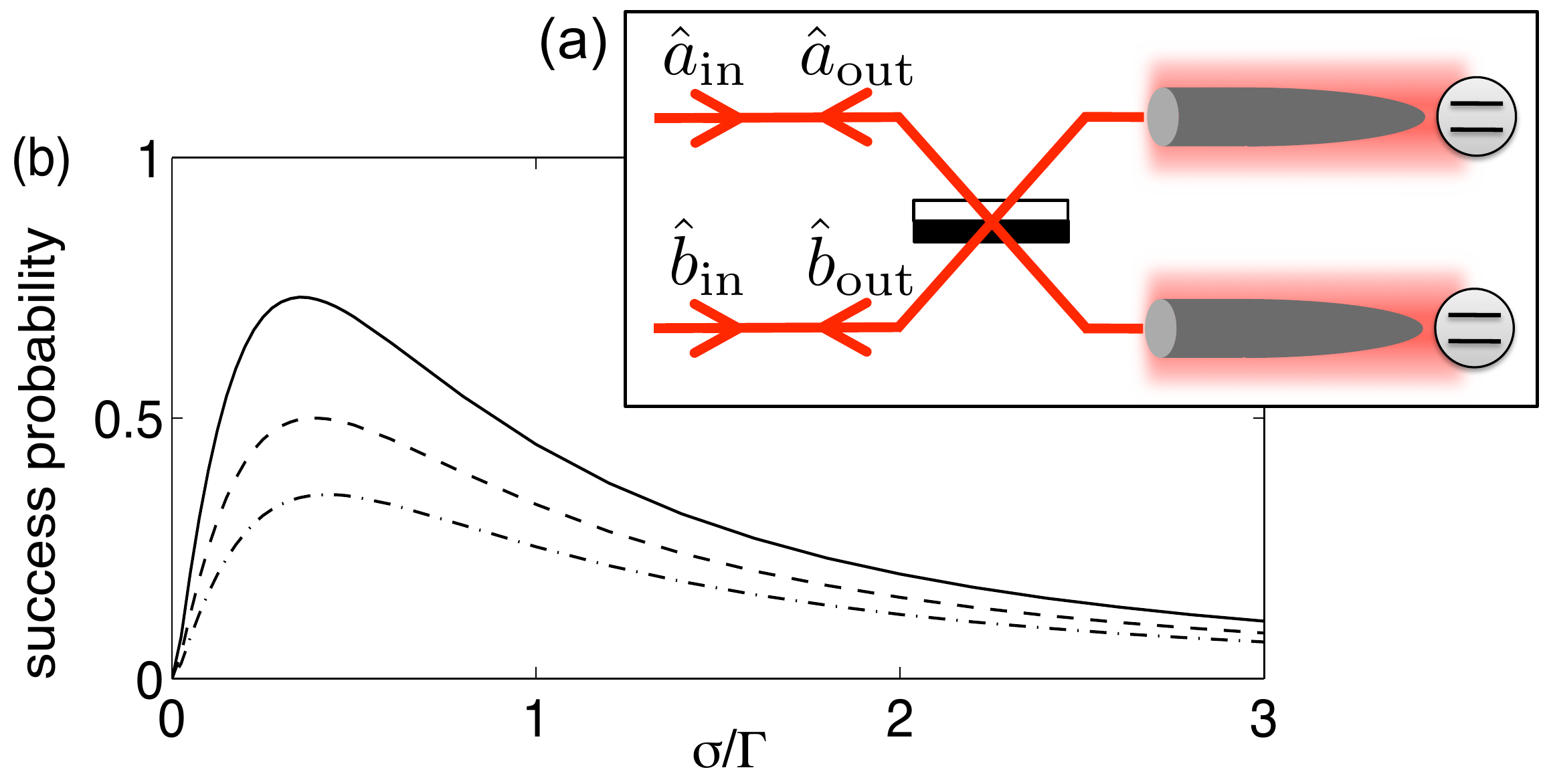}
\caption{\label{fig-nlbs1}
(a) A photon sorter based on a 1D waveguide endcoupled to a single 
emitter. A single photon will always leave the photon sorter in mode 
$\hat a_\mathrm{out}$, while two photons are likely to leave in 
mode $\hat b_\mathrm{out}$ but never split into one photon in each arm. 
(b) The success probability of the photon sorter, i.e. the probability
that two photons are scattered to the mode $\hat b_\mathrm{out}$, as a 
function of the frequency width $\sigma$ of a Gaussian input pulse for
$\gamma/\Gamma = 0$ (solid line), $\gamma/\Gamma = 0.1$ (dashed line), and
$\gamma/\Gamma = 0.2$ (dash-doted line), respectively.
}
\end{figure}

One of the conceptually simplest extensions of linear optics is a device 
capable of non-destructively distinguishing single and two photon pulses. 
Such a photon 
sorter can be realized with only passive optical elements and 
simple two level emitters using the setup sketched in 
fig.~\ref{fig-nlbs1}.

We assume that the incoming pulse enters the interferometer in the 
upper arm labelled by $\hat a_\mathrm{in}$. {A single photon wave 
packet is split at the beam splitter} and brought to interact with the emitters, 
where it experiences the phase shift (\ref{eqn-tk}). This phase depends on the 
wavenumber $k$ but is the same in both arms of the interferometer. 
The interferometer can therefore be balanced such that a single 
photon always leaves the setup in mode $\hat a_\mathrm{out}$. 
This is different if two photons enter the setup and interact indirectly 
via the emitter in which case they can leave the interferometer in mode 
$\hat b_\mathrm{out}$ with a significant probability. To be more precise,
we consider an input state of two identical photons (\ref{eqn-2phot-in}) 
with pulse shape $f_2(k,p)  = f_1(k) f_1(p)$. The beam splitter mixes 
the modes $\hat a$ and $\hat b$ such that
$
   \hat a_k^\dagger \hat a_p^\dagger
     \rightarrow  \hat a_k^\dagger  \hat a_p^\dagger
     + 2 \hat a_k^\dagger  \hat b_p^\dagger 
     + \hat b_k^\dagger  \hat b_p^\dagger. 
$
When both photons are in the same mode, the interaction with the 
emitter introduces the strongly correlated 'bound state' 
contribution (\ref{eqn-fbound}). 
Finally, after interacting with the beam splitter once again, the
two-photon input state is transformed to
\bea
   \label{eqn-turnstile-scat}
    && \!\!\!\! \ket{\Psi_\mathrm{out}} = 
     \frac{1}{2\sqrt{2}} \int \upd k  \upd p \, f_B(k,p) \hat b_k^\dagger 
       \hat b_p^\dagger \ket{0} \\
   && \; + \frac{1}{\sqrt{2}}  \int \upd k  \upd p \, 
       \left( t_k t_p f_1(k) f_1(p) + \frac{1}{2} f_B(k,p) \right) 
      \hat a_k^\dagger \hat a_p^\dagger \ket{0}. \nn
\eea
Note that there are no mixed terms (e.g. 
$\hat a_k^\dagger \hat b_p^\dagger$), so that the two photons always
leave the interferometer in the same arm. There is a 
significant probability that the two photons leave the setup in mode
$\hat b_\mathrm{out}$, whereas a single photon always leaves {the} 
interferometer in mode $\hat a_\mathrm{out}$, such that the interferometer 
 acts as a photon sorter.

The success probability of the photon sorter, i.e. the probability 
that two photons are scattered to the mode $\hat b_\mathrm{out}$ is 
given by
\be
  p_s = \frac{1}{4} \int \upd k  \upd p \, | f_B(k,p) |^2.
  \label{eqn-ptsuccess}
\ee
The success probability  is shown in fig.~\ref{fig-nlbs1} (b) for a Gaussian input 
pulse $f_1(k) \sim \exp( - (ck-\hbar \omega_0)^2 / 4 \sigma^2 )$ 
as a function of the frequency width $\sigma$. One observes that 
the efficiency of the photon sorter strongly depends on the pulse 
shape of the incident photons. They must be resonant to the atomic 
transition and the frequency width $\sigma$ should therefore not be 
too large. On the other hand the photons should be tightly localized 
in real space as they only interact when they are at the same position.
Thus an optimum value of the efficiency is
found for intermediate values of $\sigma$.

Regardless of the success probability the photon sorter 
can provide insight into the nature of the incoming light pulse. 
If for instance a conventional photo detector detects the output 
in mode $\hat b_\mathrm{out}$, the intensity of that measurement 
directly reflects the two photon contribution in the pulse. 
In addition, the success probability can be increased 
in {an} array of concatenated devices by feeding the output mode 
$\hat a_\mathrm{out}$ of one sorter to the next one.
For example, an array of five photon sorter increases the
success probability to 96 \% for $\sigma/\Gamma = 0.36$
and $\gamma=0$.

\begin{figure}[tb]
\centering
\includegraphics[width=8.2cm, angle=0]{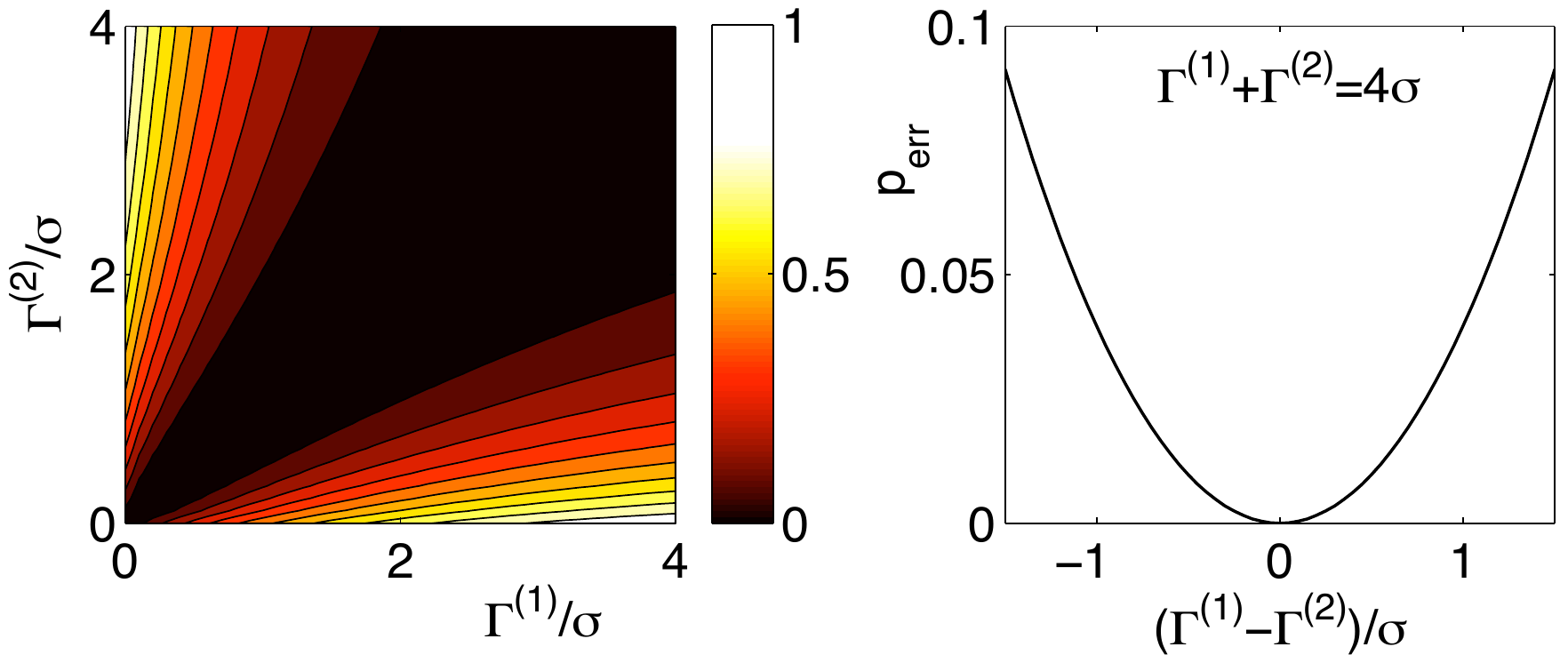}
\caption{\label{fig-sorterr}
Error probability (\ref{eqn-sorterr1}) of a photon sorter
due to a variation of the coupling rate $\Gamma^{(1,2)}$
in the two arms of the interferometer for a Gaussian
input pulse with frequency width $\sigma$.
}
\end{figure}

Coupling losses or photon absorption may reduce the success 
probability, but will not cause an incorrect routing of a single 
photon. However, such a more crucial error can occur if the 
properties of the emitters in the two interferometer arms are 
significantly different. The probability that a single photon is 
spuriously transmitted to the mode $\hat b_\mathrm{out}$ is 
then given by
\be
    p_{\rm err} = \frac{1}{4} \int dk \, | f_1(k) 
       (t_k^{(1)} - t_k^{(2)}) |^2  ,
     \label{eqn-sorterr1}  
\ee
where the superscripts $(1),(2)$ label the transmission amplitude
(\ref{eqn-tk}) for the two arms or the interferometer. This probability
is plotted in fig.~\ref{fig-sorterr} for the case of two 
different coupling rates $\Gamma^{(1,2)}$ and $\gamma=0$.
The error probability $p_{\rm err}$ vanishes quadratically with 
the difference $\Gamma^{(1)} -\Gamma^{(2)}$, such that small 
variations in the coupling rates have only weak effects.

\section{Quantum Non-Demolition Detector}

\begin{figure}[tb]
\centering
\includegraphics[width=7.5cm, angle=0]{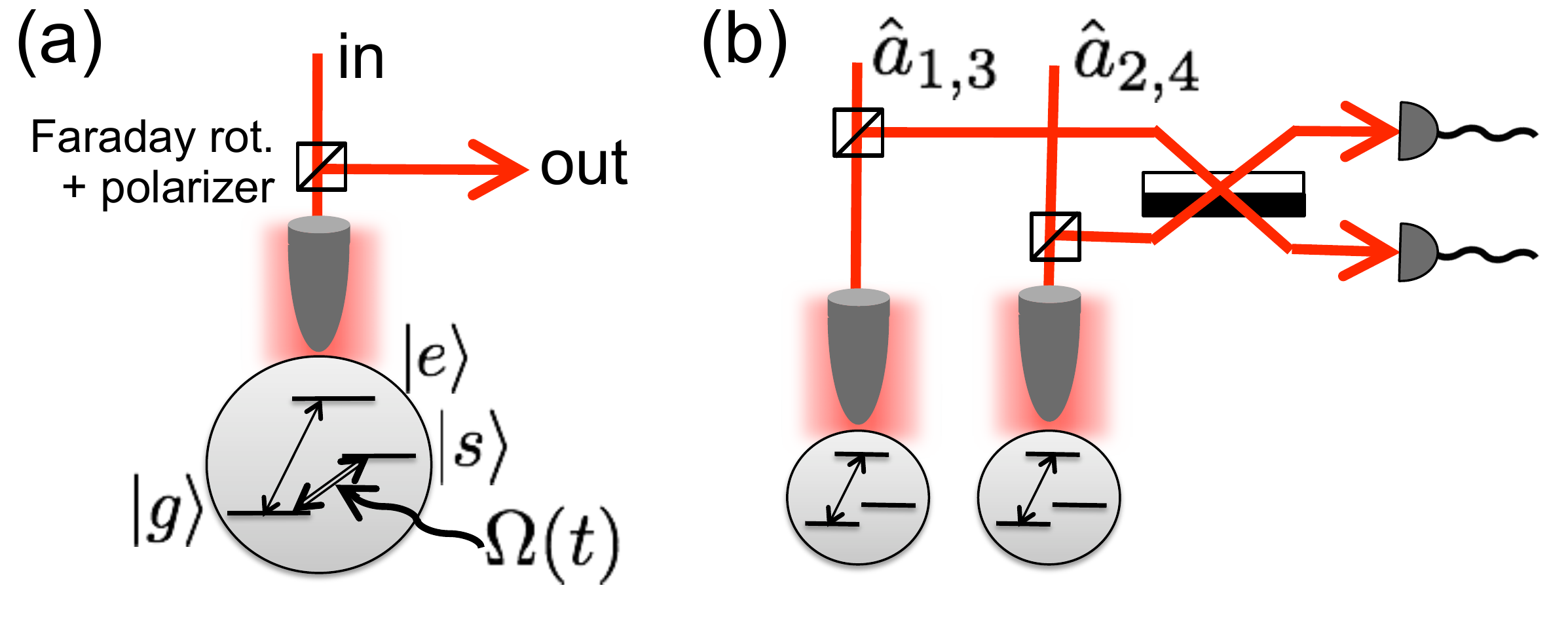}
\includegraphics[width=8.2cm, angle=0]{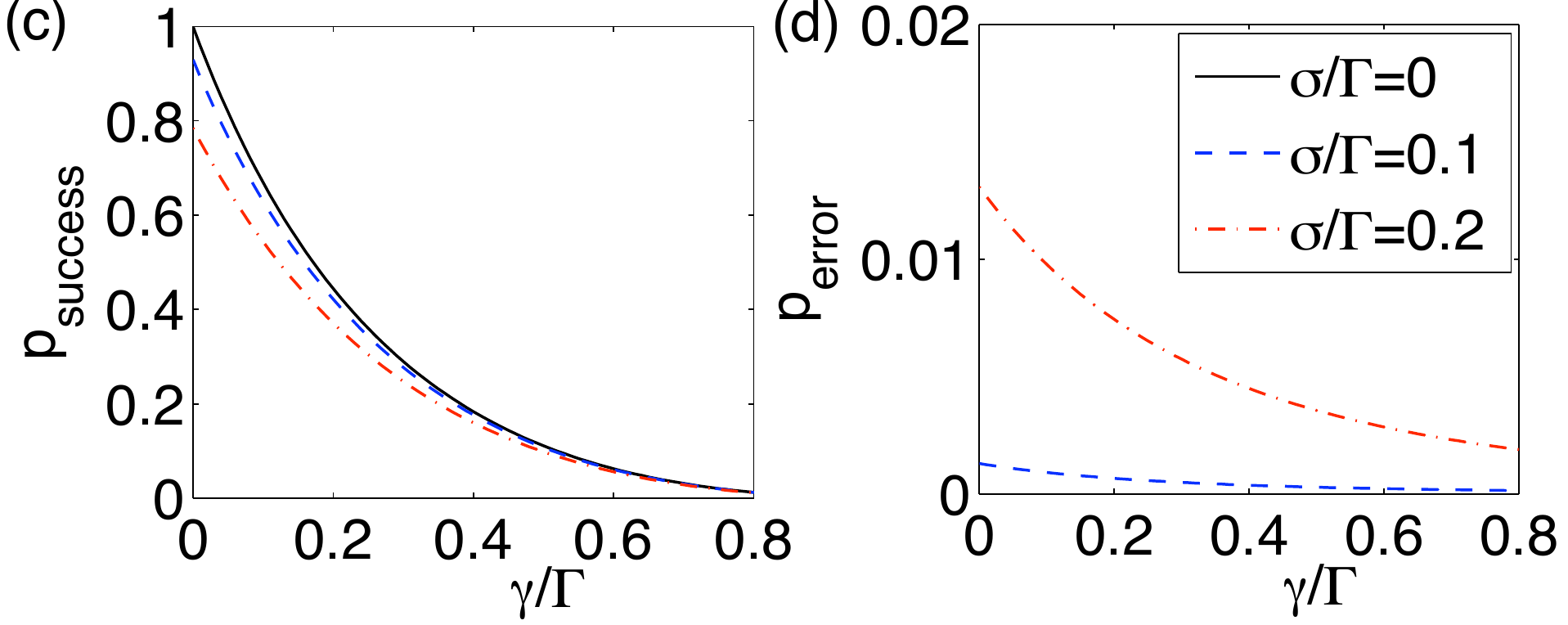}
\caption{\label{fig-dksetup}
(a) A QND photo detector based on {a three-level emitter} strongly 
coupled to a one-dimensional waveguide. A Faraday rotator combined 
with a polarizer separates the incident and the reflected mode into 
two spatially separated modes.
(b) Integration of the QND photo detector in a Bell state analyzer.
The control photon (modes $\hat a_{1,2}$) passes the 
setup before the target photon (modes $\hat a_{3,4}$).
(c,d) The success and error probabilities
as a function of the scaled decay rate $\gamma/\Gamma$ for a
Gaussian input pulse for different values of the width $\sigma$.
}
\end{figure}

To realize more advanced photo detection schemes we shall now 
consider three-level emitters as also discussed in 
Refs.~{\cite{Duan04,Chan07b,Bona10}}. 
Generalizing these approaches,  we now describe how to construct a QND 
photo detector using the setup shown in fig.~\ref{fig-dksetup} (a). 
A three-level emitter is prepared in a coherent superposition of the 
ground state $\ket{g}$ and a metastable state $\ket{s}$, which does
not couple to the waveguide,
\bea
    \ket{g} \longrightarrow \alpha \ket{g} + \beta  \ket{s}, 
    \quad \mbox{and} \quad
    \ket{s} \longrightarrow - \beta \ket{g} + \alpha \ket{s},
    \label{eqn-3dk-control}
\eea
with $\beta = \sqrt{1-\alpha^2}$. A passing resonant photon then 
introduces a phase shift if and only if the emitter is in state $\ket{g}$. 
In particular the transmission amplitude (\ref{eqn-tk}) on resonance is 
given by $t_0 = (\gamma - \Gamma)/(\gamma + \Gamma)$.
Then one applies another classical control pulse which inverts 
the transformation (\ref{eqn-3dk-control}).  The complete procedure 
thus realizes the mapping
\bea
   \mbox{1 photon:} && \ket{g} \rightarrow 
    (\beta^2  + t_k \alpha^2)  \ket{g}
     + \alpha \beta (1-t_k)  \ket{s}, \nn \\
   \mbox{0 photons:} && \ket{g} \rightarrow \ket{g}.
   \label{eqn-3dk-mapping}
\eea
If the state of the emitter is now measured to be $\ket{s}$, e.g.
by measuring the phase shift imprinted on an incident classical 
laser beam afterwards,
this unambiguously reveals the presence of a single photon. Unlike 
conventional photo detection this does not disturb the photon, i.e. the 
scheme realizes a QND measurement. Furthermore if no photon is 
present, the emitter returns deterministically to the initial state $\ket{g}$, 
i.e. there are no dark counts. The optimal detector efficiency is reached 
for $\alpha=1/\sqrt{2}$ and is given by $\Gamma/(\gamma+\Gamma)$,
which approaches unity when $\Gamma \gg \gamma$.
Using the microwave system of ref.~\cite{Asta10}, a success probability of 
$\approx 97 \%$ can be realized. 
To achieve a detector efficiency of 90\% for an optical photon, a 
modestly strong coupling with a Purcell factor of $\Gamma/\gamma=10$ 
would be required.
A different choice of $\alpha$ can be advantageous in different
contexts, in particular for the Bell state analyzer discussed below. 

\section{Bell state analyzer}
The photo detection schemes discussed above may be used in a variety 
of contexts where the measurement of more involved properties of light 
is required. A particularly important application is the design of an optical 
Bell state analyzer (BSA), which distinguishes the four Bell states 
\begin{align}
   \ket{\phi^{\pm}} &=  \frac{1}{2} \int \upd k  \upd p \, f_1(k) f_1(p)
    \left( \hat a_{1k}^\dagger \hat a_{3p}^\dagger 
     \pm \hat a_{2k}^\dagger \hat a_{4p}^\dagger  \right) \ket{0} \nn \\
   \ket{\psi^{\pm}}  &= \frac{1}{2} \int \upd k  \upd p \,  f_1(k) f_1(p)
    \left( \hat a_{1k}^\dagger \hat a_{4p}^\dagger 
     \pm \hat a_{2k}^\dagger \hat a_{3p}^\dagger  \right) \ket{0},
     \label{eqn-bellstates}
\end{align}
where the subscripts $1-4$ refer to four different photonic modes. 
In principle, a BSA  can be achieved directly from the scheme for 
photonic quantum gates in cavity QED \cite{Duan04}.
Such setups, however, often require rapid switching of the optical 
path and delay lines for photons, which is experimentally unfavorable. 
Here, we consider a modified version of ref.~\cite{Duan04}, which
avoids these elements and integrate it into a BSA. 
This setup shall be efficient and error-proof even for an imperfect
coupling, i.e $\gamma \neq 0$, so that it cannot 
give a wrong measurement result.

For resonant input photons, the QND photo detector introduced above 
is sufficient to realize a simple error-proof BSA using the setup shown 
in fig.~\ref{fig-dksetup} (b). We assume that the logical state of 
both control and target photon are encoded into two spatial 
modes. The control photon (modes $\hat a_{1,2}$) passes the 
setup well before the target photon (modes $\hat a_{3,4}$).
As described in the previous section, classical control pulses are 
applied to the emitters before and after the passage of each photon.
Thus both the control and the target photon switch the internal 
state of the respective emitter as described in 
eqn.~(\ref{eqn-3dk-mapping}). For the Bell states $\ket{\phi^\pm}$, 
both photons pass the same arm of the setup subsequently. 
The emitter coupled to this arm is transferred from state 
$\ket{g}$ to $\ket{s}$ and back to state $\ket{g}$ after interacting 
with the control and target photon, respectively. The other 
emitter always remains in the internal state $\ket{g}$.
On the contrary, the two photons pass through different arms 
of the interferometer for the Bell states $\ket{\psi^\pm}$, so 
that both emitters are transferred to the state $\ket{s}$. A 
measurement of the internal state of the emitters thus allows 
to distinguish between the subspaces spanned by $\ket{\phi^\pm}$ 
(atomic state $\ket{gg}$) on the one hand and $\ket{\psi^\pm}$ 
(atomic state $\ket{ss}$) on the other hand. 
Whether it is the plus or  the minus sign is revealed by the coincidence 
pattern of detectors placed after a beamsplitter mixing the modes 
1 and 2 as well as 3 and 4. For the plus (minus) states the photons 
will be detected in the modes 1+3 or 2+4 (1+4 or 2+3).

Taking into account photon loss, the success probability of 
this BSA is given by $p_\mathrm{success} = \eta^2 |t_0|^2 = 
\eta^2(\gamma - \Gamma)^2/(\gamma + \Gamma)^2$, 
where $\eta$ is the efficiency of the final photo detectors
(assuming $\alpha=1/1-t_0$, see below).
With non-resonant input, the probability for a successful Bell 
measurement including any transmission losses is given by
\be
   p_\mathrm{success} = \frac{ \eta ^2 |t_0|^2}{|1-t_0|^4} 
   \int \upd k  \upd p \, | f_2(k,p) (1-t_k) (1-t_p) |^2
\ee
regardless of which of the Bell states is incident. This result (with $\eta=1$) 
is plotted in fig.~\ref{fig-dksetup} (c) as a function of the loss rate
$\gamma/\Gamma$ for Gaussian input pulses.
One finds that a Purcell factor 
of $\Gamma/\gamma \approx 5.8$ is sufficient to exceed
the $\eta^2\times50$ \% limit of linear optics.

The present setup is, however, not strictly
error-proof if the input photons are not completely resonant. 
While the measurement result $\ket{ss}$ leads to an unambiguous 
Bell state measurement, the result $\ket{gg}$ does not. It can be 
almost certainly attributed to the subspace spanned by $\ket{\phi^\pm}$, 
but there is a small probability that it has been triggered by the states 
$\ket{\psi^\pm}$. This error can, however, be suppressed to a large
extend by choosing a rotation angle of $\alpha = 1/(1-t_0)$. In this case the residual probability to obtain an erroneous measurement result for the input state $\ket{\psi^\pm}$ is given by
\be
   p_\mathrm{error} = \frac{\eta^2}{|1-t_0|^4} \int \upd k  \upd p \, 
     | f_2(k,p) (t_0-t_k) (t_0-t_p) |^2.
\ee
For a Gaussian wavepacket, $p_\mathrm{error}$ vanishes as
$\sigma^4/(\gamma+\Gamma)^4$ for $\sigma/(\gamma+\Gamma )\rightarrow 0$. 
It thus remains small also for a non-monochromatic input photon 
as shown in fig.~\ref{fig-dksetup} (d).

In order to realize a fully error-proof BSA, 
one needs another measurement stage, which unambiguously 
detects $\ket{\phi^\pm}$. In principle this can be realized by 
exchanging the modes $\hat a_3$ and $\hat a_4$ and 
then repeating the above scheme. This would, however, require  
rapid switching of the optical path between control and target 
photon.

\begin{figure}[tb]
\centering
\includegraphics[width=8cm, angle=0]{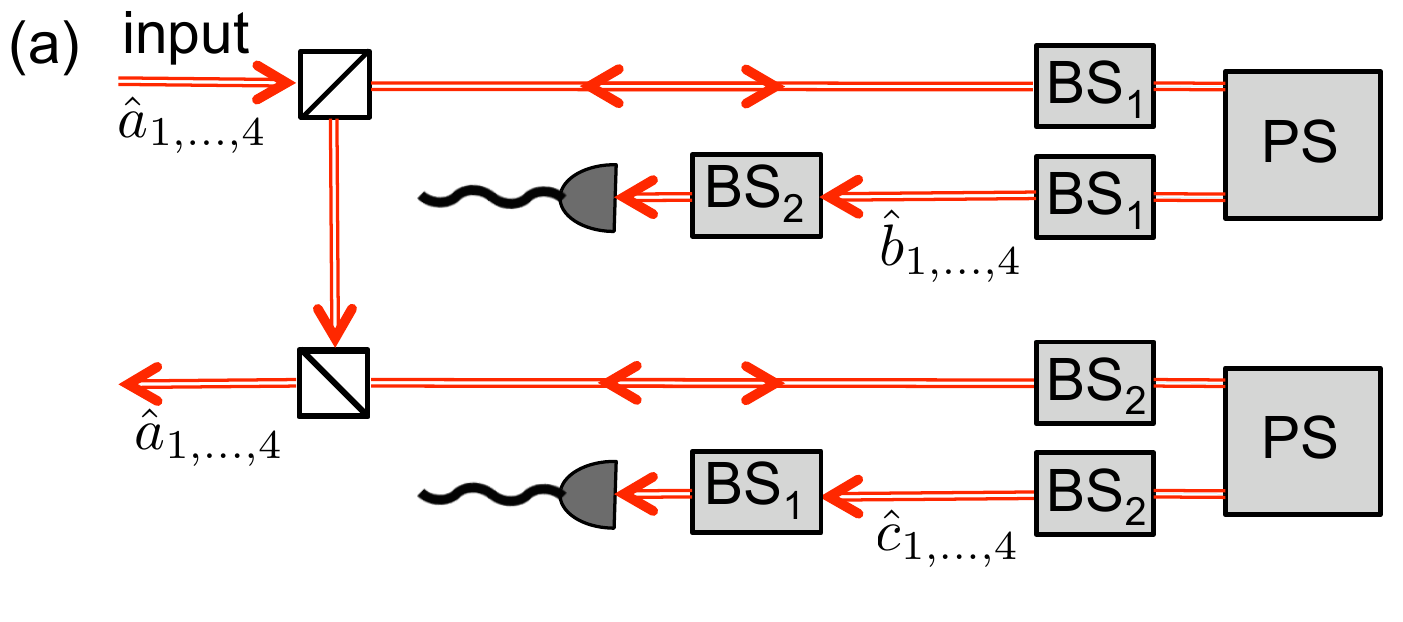}
\includegraphics[width=8cm, angle=0]{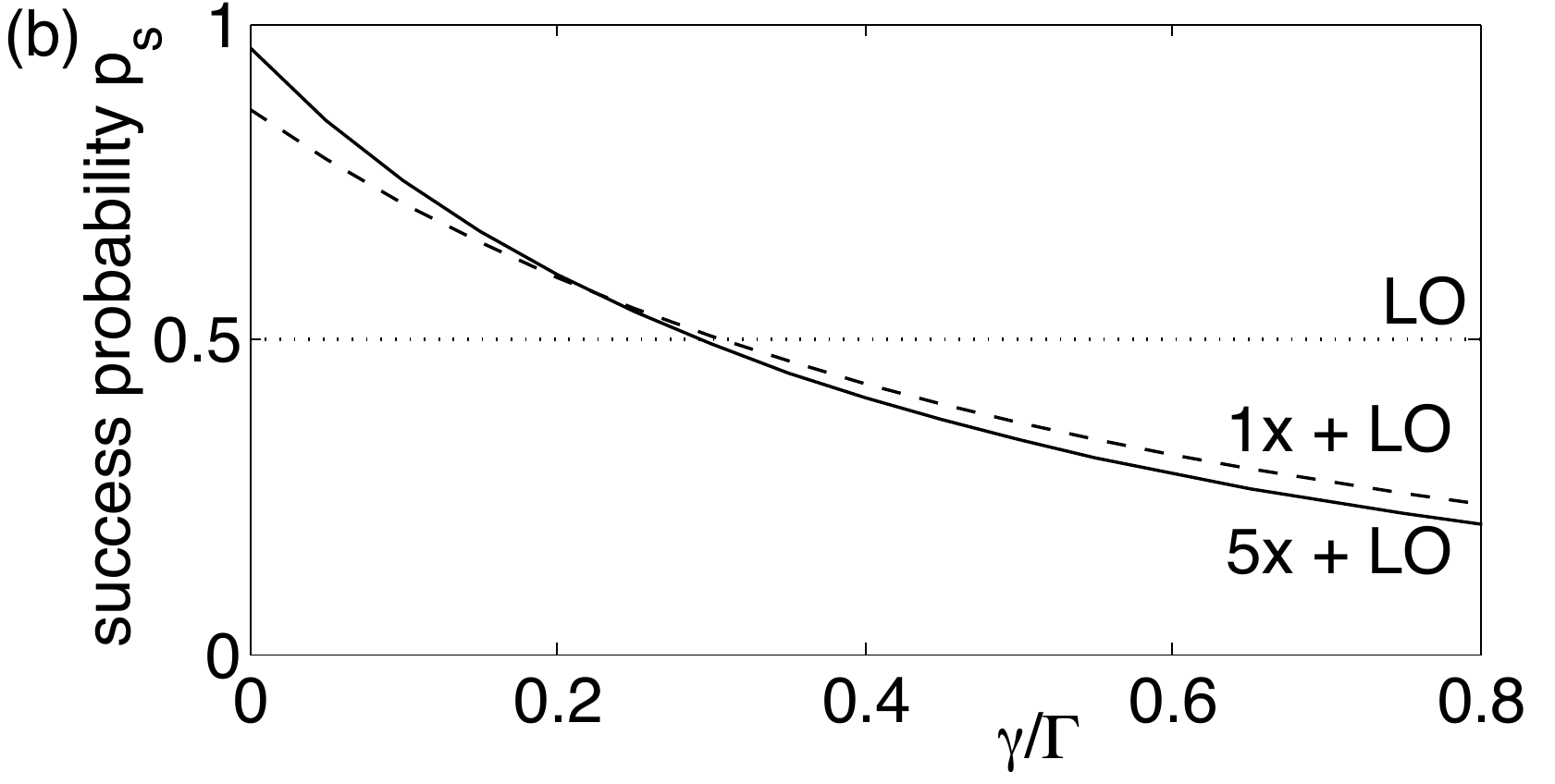}
\caption{\label{fig-setup}
(a) Setup of a BSA composed of two photon sorters (PS) 
and linear beam splitter arrays ($\mathrm{BS}_1$ and 
$\mathrm{BS}_2$, see text). The crossed squares are
Faraday rotators and polarizers separating incoming and 
reflected modes. Each of the arrows represents four modes 
carrying the Bell state.
(b) Total success probability of an array of $n=1$ (1x + LO) and 
$n=5$ (5x+LO) error-proof BSAs,  plus a linear optical 
BSA at the end of the array as a function of the ratio $\gamma/\Gamma$ 
assuming a Gaussian pulse shape with frequency width
$\sigma/\Gamma = 0.36$.
The success probability of a linear optical BSA (LO) is plotted
for comparison.
}
\end{figure}  

\section{A passive Bell state analyzer}
The detection scheme described above realizes a perfect
Bell state measurement in the ideal case, but requires a
three-level system and an active control of the emitter.
As we will now show, a fully passive, error-proof BSA may in fact be 
constructed using the  photon sorters introduced above. The setup 
to perform these operations is summarized in fig.~\ref{fig-setup} (a).
Assume that the four optical modes containing the Bell state in 
eq.~(\ref{eqn-bellstates}) are incident on a beam splitter array mixing 
the modes 1 and 4 as well as 2 and 3 (denoted by $\mathrm{BS}_1$ in 
fig.~\ref{fig-setup}).
The states $\ket{\psi^\pm}$ are thereby mapped 
onto $ \sim (\hat a_{1}^{\dagger2} - \hat a_{4}^{\dagger2} 
   \pm \hat a_{2}^{\dagger2}  \mp \hat a_{3}^{\dagger2} ) \ket{0} $, 
suppressing the pulse shape for simplicity.
The two photons are always located in the same mode for
the states $\ket{\psi^\pm}$, whereas they are always located in two
different modes for the states $\ket{\phi^\pm}$. If each of the modes
is now incident on the photon sorter
 introduced above, the states $\ket{\psi^\pm}$ are separated
to the modes $\hat b_{1,\ldots,4}$ with a significant probability. It is 
then possible to distinguish between $\ket{\psi^+}$ and $\ket{\psi^-}$ 
with linear optics and conventional photodetectors only, giving rise to an unambigious 
Bell state measurement. If no photon is detected we can simply go on 
with the two photons in the modes $\hat a_{1,\ldots,4}$. In order to
detect also the Bell states $\ket{\phi^\pm}$, one undoes the effect
of the first beam splitter array $\mathrm{BS}_1$ and then mixes the
modes 1 and 3 as well as 2 and 4 instead (denoted by $\mathrm{BS}_2$ in 
fig.~\ref{fig-setup}). Now it's the states $\ket{\phi^\pm}$, for which the 
two photons are  located in the same mode, these are thus
separated to the modes $\hat c_{1,\ldots,4}$ by the photon sorter 
and subsequently detected.  The photons
are either measured in the two detector arrays projecting 
unambiguously onto one of the Bell states or leave the BSA
in the modes $\hat a_{1,\ldots,4}$. In the latter case another 
measurement can be attempted.

The proposed BSA works probabilistically -- with a non-vanishing
probability the photons are not detected but just transmitted   
through the complete setup. The success probability is given 
by the probability of two photons to be scattered to the modes 
$\hat b_{1,\ldots,4}$ or $\hat c_{1,\ldots,4}$, respectively,
and thus given by the success probability of the photon sorter,
which is given in eq.~(\ref{eqn-ptsuccess}) and plotted 
in fig.~\ref{fig-nlbs1} (b).
If the detection fails and the photons are transmitted,
one can just repeat all operations. However, in actual experiments 
photon losses are inevitable and the coupling of the emitter to the 
one-dimensional waveguide is not perfect.
Photon loss only leads to an inconclusive measurement result and  
fig.~\ref{fig-setup} (b) shows the resulting success probability of an array of 
1 (dashed line) and 5 (solid line) concatenated BSAs,
as a function of the ratio $\gamma/\Gamma$. After passing this 
array, we assume that the remaining modes are detected with a linear optical BSA.
As shown in the figure already a modest Purcell factor of 
$\Gamma/\gamma \approx 3.3$
is sufficient to exceed the  50 \% limit of linear optics.

\section{Conclusion}

We have shown that the coupling of single emitters to one-dimensional 
waveguides opens up new possibilities for number resolving, non-demolition 
photo detection. We have explicitly shown how to construct  photon sorters 
and QND detectors, and that these systems can be used for efficient Bell 
state analysis.   Most importantly the devices are error-proof in the sense 
that imperfect coupling only leads inconclusive and not wrong results. As 
a consequence the devices work with modest coupling efficiencies, which 
are well within reach of current experiments.


\acknowledgments
Financial support by the German Research Foundation 
(DFG grant WI 3415/1), the Danish National Research Foundation and 
the Villum Kann Rasmussen foundation is gratefully acknowledged. 
Work at Harvard was supported by NSF, CUA,  DARPA and Packard Foundation.

\end{document}